\documentclass[conference]{IEEEtran}
\IEEEoverridecommandlockouts
\usepackage{cite}
\usepackage{amsmath,amssymb,amsfonts}
\usepackage{algorithmic}
\usepackage{graphicx}
\usepackage{textcomp}
\usepackage{changepage}
\usepackage{xcolor}
\usepackage{array}
\usepackage{booktabs} 
\usepackage{newtxtext} 
\usepackage{color}
\usepackage{hyperref}
\usepackage{url}
\usepackage{tabularx}
\usepackage[table]{xcolor} 
\usepackage{float}
\usepackage{multirow}

\usepackage{longtable}
\usepackage{graphicx} 
\usepackage{lscape} 
\usepackage{lscape} 

\usepackage{listings}
\usepackage{cleveref}
\usepackage{mathrsfs} 
\usepackage{graphicx} 
\usepackage{subcaption} 
\usepackage{makecell}
\usepackage{booktabs}
\usepackage{pifont}
\usepackage{adjustbox}  
\usepackage{booktabs} 
\renewcommand{\arraystretch}{1.2}
\def\BibTeX{{\rm B\kern-.05em{\sc i\kern-.025em b}\kern-.08em
    T\kern-.1667em\lower.7ex\hbox{E}\kern-.125emX}}
    
\begin{document}

\title{From Indiscriminate to Targeted: Functionally Critical Signal-Driven Assertion Generation using LLMs for Efficient RTL Verification
}

\author{
\IEEEauthorblockN{
Yonghao Wang\textsuperscript{1},
Hongqin Lyu\textsuperscript{1,2},
Boling Chen\textsuperscript{3}, 
Jiaxin Zhou\textsuperscript{4}, 
Minyang Bao\textsuperscript{1},
Wenchao Ding\textsuperscript{5},
Feng Gu\textsuperscript{6},
\\
Zhiteng Chao\textsuperscript{1,2},
Jianan Mu\textsuperscript{1,2},
Kan Shi\textsuperscript{1,2},
Tiancheng
Wang\textsuperscript{1}
and
Huawei Li\textsuperscript{1,2}}
\IEEEauthorblockA{\textsuperscript{1}State Key Lab of Processors, Institute of Computing Technology, CAS, Beijing, China}
\IEEEauthorblockA{\textsuperscript{2}University of Chinese Academy of Sciences, Beijing, China}
\IEEEauthorblockA{\textsuperscript{3}Beijing University of Posts and Telecommunications, China; \textsuperscript{4}Beijing Normal University, China;}
\IEEEauthorblockA{\textsuperscript{5}Tencent, China; \textsuperscript{6}The Chinese University of Hong Kong, China}
\IEEEauthorblockA{\{wangyonghao22, lvhongqin24\}@mails.ucas.ac.cn,
\{chaozhiteng, wangtiancheng, lihuawei\}@ict.ac.cn}
}

\maketitle      

\begin{abstract}
Functional verification has become the most time-consuming phase in IC development, and Assertion-Based Verification (ABV) is key to reducing debugging time. However, existing LLM-based assertion generation methods typically pursue indiscriminate verification, aiming for maximal coverage without considering signal criticality, whereas industrial practice demands maximizing coverage with minimal verification cost. Consequently, identifying signals that have the greatest impact on design functionality and error propagation—enabling a shift from indiscriminate to targeted verification—remains a key challenge. To address this, we propose AgileAssert, a critical signal-driven assertion generation framework that constructs RTL semantic graphs and identifies the top-K critical signals via a hybrid scoring and selection mechanism, followed by structure-aware RTL slicing to provide the LLM with precise targets and contextual information, thereby guiding LLMs to generate tightly constrained targeted assertions for efficient verification. Evaluated on block- and CPU-level designs, with an average 66.68\% reduction in assertions, our approach outperforms three existing SOTA methods, and significantly improving coverage metrics while reducing input token consumption by 64\%. In mutation testing, when our approach surpasses existing methods in error detection rate, the average number of assertions used decreases by 72.74\%.
\end{abstract}


\section{Introduction}

Functional verification plays a crucial role in the realm of integrated circuit (IC) design. Verification engineers meticulously evaluate whether the register-transfer level (RTL) code adheres to the prescribed architectural specifications. Assertion-Based Verification (ABV) has garnered widespread adoption in RTL design, primarily due to its efficacy in enhancing visibility and substantially reducing the time required for simulation debugging, with reductions of up to 50\% \cite{1, 2}. Within the ABV framework, the significance of high-quality SystemVerilog assertions (SVA) for formal property verification (FPV) cannot be overstated \cite{3, 4}. 

\begin{figure}[h]
\centering
\includegraphics[width=1\linewidth]{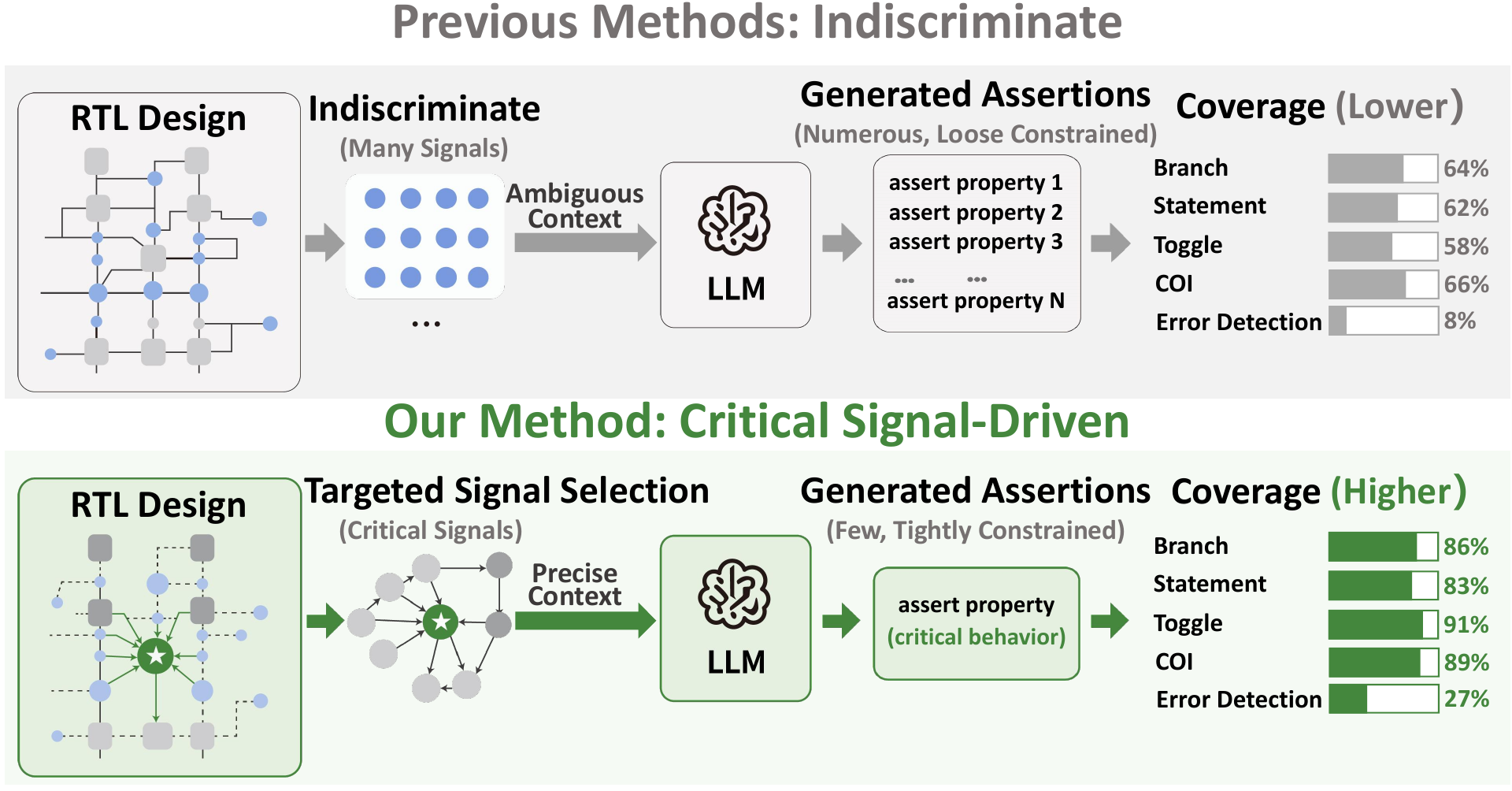}
\caption{Efficient RTL verification through critical signal-driven assertion generation.}
\label{fig:intro}
\end{figure}

With the growing complexity of computer architecture and hardware designs, functional verification has become time-intensive throughout the entire chip development cycle. A recent study indicates verification accounts for up to 70\% of development efforts \cite{5}, making it essential to find appropriate techniques to accelerate this process. Consequently, efficient SVA generation has become crucial. 

Traditional assertion generation methods \cite{b1, b2, b3, b4} primarily rely on mining simulation waveforms or execution traces, combined with machine learning models for pattern induction, the generated assertions are often trivial and semantically fragmented, making it difficult to capture higher-level behavioral semantics. Recent advances in LLMs have shown great potential to revolutionize this process, offering innovative ways to automate assertion generation. These methods fall into two primary approaches: pre-RTL SVA generation, which directly converts natural language specifications into assertions \cite{6,7,8}, and methods that integrate both specifications and RTL code to capture finer functional and structural details \cite{9, 10, 11, 12, 27, 31}.

However, despite the promising progress, both categories of existing approaches exhibit notable limitations when applied to large-scale industrial designs. For pre-RTL SVA generation methods that rely solely on natural language specifications, the primary bottleneck lies in the inherent abstraction and incompleteness of specifications. In practice, specifications often describe only high-level functionality and top-level interfaces, while omitting many internal signals and corner-case behaviors. As a result, assertions generated from specifications alone tend to focus on coarse-grained properties, making it difficult to achieve high functional coverage. This limitation becomes even more pronounced as circuit complexity increases.

On the other hand, approaches that incorporate RTL code alongside specifications aim to address this issue by leveraging structural and behavioral information from the design. While such methods can, in principle, generate more precise and comprehensive assertions, they face significant scalability challenges. Modern IC designs often consist of millions of lines of RTL code, which can either exceed the input token limits of current LLMs or produce overly long contexts that make it difficult for the model to focus on critical behaviors \cite{13}. Feeding the entire design into the model is therefore impractical, and unwarranted truncation or sampling strategies inevitably lead to severe loss of critical information. Moreover, processing large-scale RTL incurs substantial computational overhead, making these approaches inefficient and difficult to deploy in real-world verification flows.

More fundamentally, previous works implicitly pursue a form of indiscriminate or uninformed verification, relying on the generation of a large number of assertions in an attempt to achieve comprehensive coverage. However, this strategy provides the LLM with a highly ambiguous context and lacks a clear functional focus, as the verification is not explicitly guided toward specific, high-impact functionalities. As a result, the generated assertions are loosely constrained and often of low quality. Moreover, as circuit scales continue to grow, achieving full coverage through sheer assertion volume becomes increasingly infeasible in terms of time and computational resources. In industrial practice, the key challenge lies not in exhaustive verification, but in efficiently identifying critical design components and maximizing coverage with minimal verification effort \cite{14, 30}.

To address these challenges, we propose AgileAssert, a novel assertion generation framework that shifts focus from indiscriminate to targeted verification. Our method first designs a dedicated RTL parser to classify signals and extract a semantic graph with data propagation types. By extracting four features that capture the greatest impact on design functionality and error propagation, it identifies critical signals and further refines them through structure-aware RTL slicing. This provides the LLM with precise targets and contextual information, thereby fully leveraging its powerful semantic understanding to generate tightly constrained assertions for efficient verification, as shown in Fig.\ref{fig:intro}.

The contributions of this paper are summarized as follows:
\begin{enumerate}[]
\item We propose a hybrid signal scoring framework to identify critical signals for targeted verification, enabling efficient functional checking by focusing on critical signals that impact design functionality and error propagation mostly.
\item We develop a structure-aware RTL slicing method to extract precise code fragments centered on critical signals, mitigating token overflow and context noise, and enabling scalable verification of large-scale designs while reducing computational cost.
\item Experiments on block- and CPU-level designs show that, with an average 66.68\% reduction in assertions, our framework achieves improvements in branch, statement and toggle coverage by 14.13\%, 12.96\% and 11.99\%, respectively, and improves Cone of Influence (COI) coverage by 21.27\%, while reducing input token consumption by 64\%. In mutation testing, when our approach surpasses existing methods in error detection rate, the average number of assertions used decreases by 72.74\%.
\end{enumerate}

\section{Background}

\subsection{Assertion Generation Based on Traditional Methods}

The early work of automated hardware assertion generation was significantly advanced by Vasudevan et al. \cite{b3}, who proposed GoldMine, a framework synergistically combining data mining and static analysis to extract propositional and temporal assertions from RTL simulation traces. Malburg et al. \cite{b4} introduced a dynamic dependency graph-based approach that generates SystemVerilog properties from limited simulation runs through symbolic value abstraction and model checking. Subsequently, Germiniani and Pravadelli \cite{b1} presented HARM, a highly customizable hint-based assertion miner employing user-defined LTL templates and entropy-based decision trees to efficiently mine temporal assertions from execution traces. However, these traditional methods are fundamentally constrained by their reliance on pre-defined templates, local static analysis, or trace-specific mining, often producing simple propositional assertions with limited semantic expressiveness and failing to capture complex cross-module behaviors or high-level architectural intent.

\subsection{Assertion Generation Based on LLMs}

The early work of automated hardware assertion generation proposed by Rahul Kande et al. \cite{10}, who pioneered the application of LLMs to generate assertions. AssertLLM by Fang et al. \cite{6} was a milestone, handling full specification files to produce detailed SVAs for all architectural signals. Bai et al. \cite{9} contributed AssertionForge, which built a unified Knowledge Graph (KG) from both natural language specs and RTL code, linking high-level design intent with low-level implementation details. More recently, Wang et al. \cite{12} introduced DeepAssert, leveraging the powerful reasoning capabilities of LLM to derive module-level specifications by extracting reliable information from the RTL to be verified. This further generates fine-grained module-level assertions, achieving more precise constraints and bug localization.

\subsection{Coverage-Driven Efficient Verification}

Improving verification efficiency under limited resources has long been a critical challenge in large-scale RTL design validation \cite{14, 30}. Traditional approaches often rely on static structural and semantic analysis of specifications and RTL code to iteratively improve assertion completeness \cite{6, 8, 9, 11}. Existing studies have explored coverage-driven methodologies to guide the verification process \cite{30, 27}, where coverage metrics are used as feedback signals to identify insufficiently verified design regions and refine verification strategies accordingly. However, they typically require generating a large number of assertions to approach satisfactory coverage, resulting in substantial computational overhead and prolonged verification cycles, and often struggle with large-scale designs.

\section{Framework of AgileAssert}

\subsection{Workflow Overview}

\begin{figure*}[h]
\centering
\includegraphics[width=0.9\linewidth]{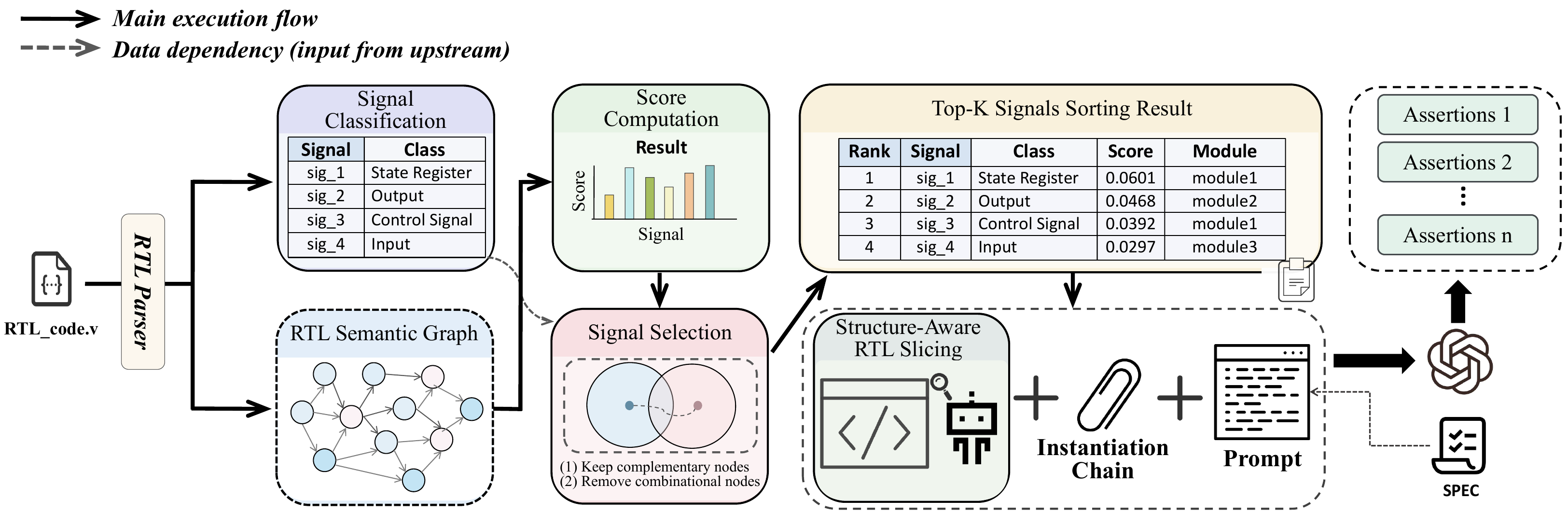}
\caption{The workflow of the AgileAssert framework. Our RTL parser constructs a semantic graph from the RTL code. Signals are then scored using multiple features and filtered through selection mechanism to produce a ranked list of signals. These signals guide structure-aware RTL slicing, which in turn directs LLM-based targeted assertion generation.}
\label{fig:framework}
\end{figure*}

This paper proposes AgileAssert, a signal-centric assertion generation framework that identifies functionally critical signals from RTL designs and generates targeted assertions accordingly. By shifting from indiscriminate verification to focused analysis, AgileAssert improves verification efficiency and scalability for large-scale designs. The overall workflow is illustrated in Fig.\ref{fig:framework}.

\subsection{Semantic Graph Extraction}

The first step in our framework involves constructing a semantic graph from RTL code, where each node represents a signal and edges encode functional dependencies. Formally, we define the graph as:
\[
G = (V, E), \quad V = \{v_1, v_2, \dots, v_n\}, \quad E \subseteq V \times V,
\]
where \(V\) is the set of signal nodes and \(E\) represents directed edges capturing functional dependencies between signals.

To enable fine-grained RTL structural analysis, we develop a custom RTL parser that processes RTL code to classify signals, extract module instantiation chains, and construct an RTL semantic graph among variables.

\subsubsection{\textbf{Signal Classification}}

Signals are initially categorized into four functional classes: 
\textit{state registers}, \textit{control signals}, \textit{output ports}, and \textit{internal signals}. \textit{State Registers} are identified as assignment targets within \texttt{always} blocks. They capture FSM states, data flow control, and general sequential logic. \textit{Control Signals} are derived from conditional expressions that govern branching and loop structures. \textit{Output Ports} are extracted directly from module declarations. \textit{Internal Signals} include all remaining signals that do not fall into the above categories.

To avoid naming ambiguity across hierarchical designs, we adopt a fully qualified naming scheme by concatenating the module name with the signal name. This ensures that signals from different modules are treated as distinct nodes even if they share identical identifiers. Nodes belonging to different modules in the graph are identified by different colors.

In addition, we analyze whether each signal is purely combinational, assigning a flag to indicate this property. Specifically, if all code fragments associated with a signal consist solely of combinational logic, the signal is marked as \texttt{True}; otherwise, it is marked as \texttt{False}.

\subsubsection{\textbf{Edge Types}}

Based on the traversal of abstract syntax tree (AST) implemented in our parser, we extract four categories of semantic edges:

\begin{itemize}
\item \textbf{Data Dependency:}
For assignment statements (including continuous, blocking, and non-blocking assignments), we establish directed edges from right-hand side (RHS) signals to left-hand side (LHS) signals, representing combinational data flow.

\item \textbf{Temporal Dependency:}  
When assignments occur within edge-triggered \texttt{always} blocks (i.e., sensitive to \texttt{posedge} or \texttt{negedge}), the corresponding dependencies are labeled as temporal. These edges capture state transitions across clock cycles and explicitly encode sequential behavior.

\item \textbf{Control Dependency:}  
For conditional constructs such as \texttt{if}, \texttt{case}, and \texttt{for} statements, we extract control signals from condition expressions and connect them to all signals assigned within the corresponding branches. This models the control-flow influence on signal updates.

\item \textbf{Module Interaction:}  
To capture inter-module connectivity, we analyze module instantiations and port bindings, adding edges between parent-module signals and submodule ports based on port directions (input/output), thereby reflecting hierarchical signal propagation and preserving the inter-module call hierarchy, which provides crucial context for subsequent signal slicing and LLM-guided assertion generation.
\end{itemize}

\subsubsection{\textbf{AST-Based Graph Extraction}}

\begin{figure}
\centering
\includegraphics[width=1\linewidth]{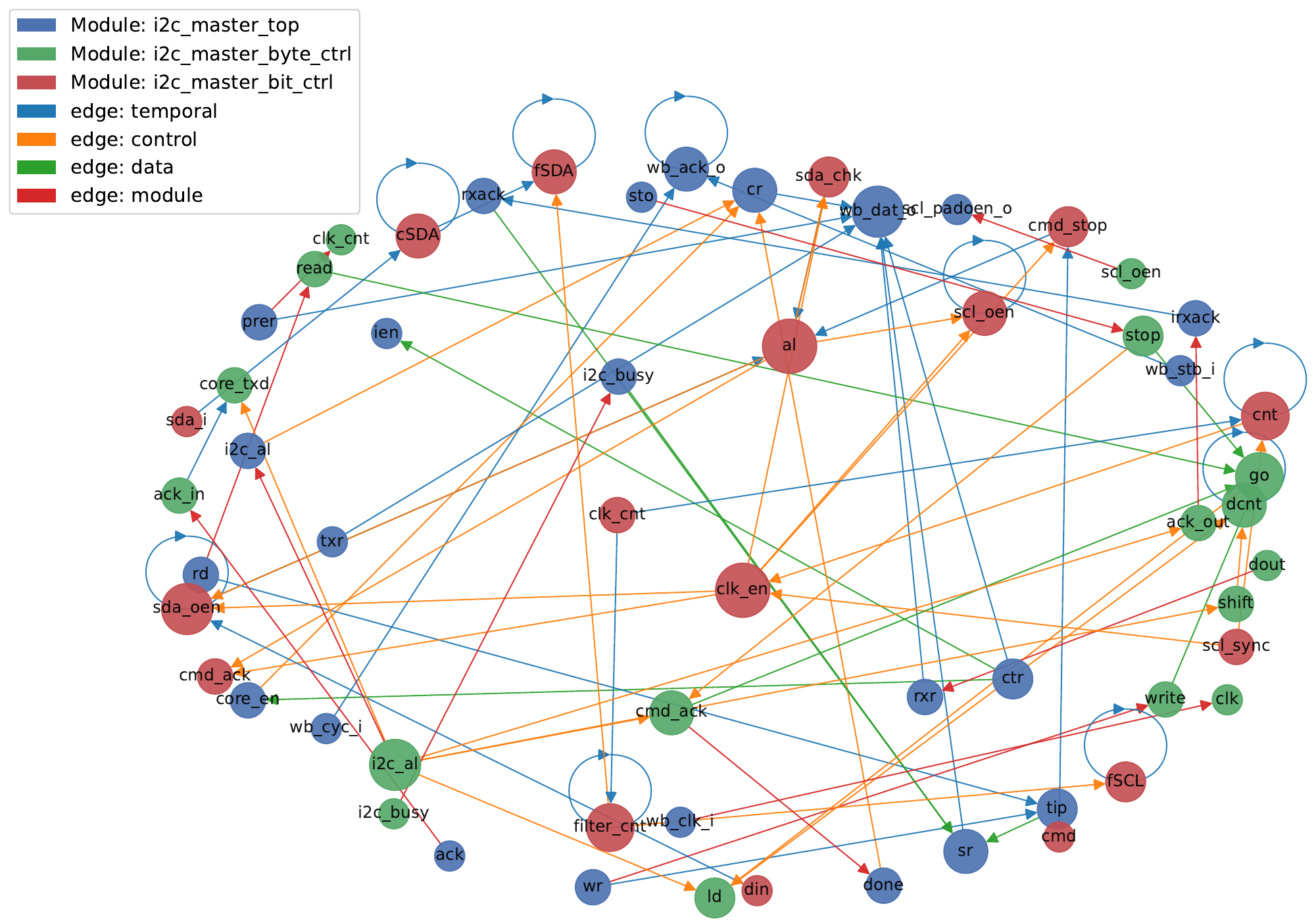}
\caption{Example of a simplified RTL semantic graph from an I2C circuit \cite{6}. The design consists of three modules.}
\label{fig:semantic_graph}
\end{figure}

Our approach performs a recursive traversal over the Verilog AST. For each statement, identifiers are extracted from both expressions and assignment targets. The parser distinguishes between different statement types (e.g., assignment, conditional, loop) and applies corresponding extraction rules to ensure accurate dependency modeling. In particular, we recursively collect LHS signals in nested control structures to correctly associate control conditions with all affected assignments. An example of an RTL semantic graph extracted from the I2C circuit is shown in Fig.\ref{fig:semantic_graph}.

\subsection{Signal Ranking}

To identify functionally critical signals for assertion generation, we propose a hybrid ranking framework that integrates structural, temporal, and control-flow characteristics. The overall process consists of three stages: noise-reduced subgraph filtering, multi-feature scoring, and redundancy-aware top-k selection.

\subsubsection{\textbf{Noise-Reduced Subgraph Filtering}}

Although the constructed RTL semantic graph captures comprehensive dependencies, not all signals contribute equally to multi-cycle functional behaviors. To focus on behaviorally dominant signals, we extract a noise-reduced subgraph by removing non-functional signals (e.g., clock, parameters, and redundant internal temporaries), along with self-loops. The resulting subgraph preserves both sequential and combinational behaviors relevant for functional analysis. By restricting analysis to the refined subgraph, we reduce noise from local combinational logic while preserving the global behavioral structure.

\subsubsection{\textbf{Multi-Feature Scoring}}

To quantify the behavioral significance of signals within the RTL semantic graph, we compute a \emph{hybrid importance score} that integrates multiple complementary structural and functional metrics. This scoring framework captures both global topological influence and signal-specific characteristics relevant to functional verification, specifically including the following features:

\paragraph{\textbf{PageRank Score}}
The \emph{PageRank score} $PR(v)$ evaluates the structural centrality of a signal within the global RTL dependency graph. It reflects how strongly a signal connects to other important nodes, indicating its role as a critical hub or mediator in information propagation. Formally, given the directed graph $G=(V,E)$, the $PR(v)$ for node $v \in V$ is computed following the canonical PageRank algorithm \cite{18}:

\begin{equation}
PR(v) = (1-\alpha) + \alpha \sum_{u \in pred(v)} \frac{PR(u)}{outdeg(u)}
\end{equation}
where $\alpha$ is the damping factor, set to the commonly used value 0.85, $pred(v)$ denotes the set of predecessors of $v$, and $outdeg(u)$ is the out-degree of node $u$. This metric highlights signals that occupy central positions in the dependency network of the design.

\paragraph{\textbf{Observability Score}}
The \emph{Observability score} measures the extent to which a signal affects other signals in the circuit through its backward reachable set. For each node $v$, we traverse the graph along incoming edges and count the number of outgoing \texttt{control} or \texttt{data} dependencies of its predecessors:

\begin{equation}
\begin{split}
Obs(v) = \sum_{p \in \text{Backward}(v)} & \; 
\bigl| \{ (p,v) \in E \mid \\
& \text{type}(p,v) \in \{\text{control}, \text{data}\} \} \bigr|
\end{split}
\end{equation}

Intuitively, signals with higher observability scores are more influential in propagating functional behaviors, as they control or impact a larger number of dependent signals.

\paragraph{\textbf{OutputBoost Score}}
The \emph{OutputBoost score} amplifies the importance of signals that serve as module outputs. Specifically, if a signal is classified as an output port, its observability value is directly incorporated as a boost:

\begin{equation}
OutBoost(v) = 
\begin{cases}
Obs(v), & \text{if } v \text{ is an output} \\
0, & \text{otherwise}
\end{cases}
\end{equation}

This mechanism prioritizes signals that contribute to externally observable behaviors, ensuring that critical endpoints in the circuit are emphasized during top-k signal selection.

\paragraph{\textbf{MuxBranch Score}}
The \emph{MuxBranch score} captures whether a signal participates in multiple branching conditions, such as case statements or multiplexers. For each signal, we count the number of outgoing control edges targeting nodes that are controlled multiple times:

\begin{equation}
\begin{split}
\text{MuxBranch}(v) = \sum_{u \in succ(v)} & \; 
\mathbb{I} \Big( |\{ (v,u) \in E \mid \\
& \text{type}(v,u) = \text{control} \}| \ge 1 \Big)
\end{split}
\end{equation}
where $succ(v)$ denotes the set of signals directly influenced by signal $v$ through outgoing edges in the RTL semantic graph, and $\mathbb{I}(\cdot)$ is the indicator function that returns 1 when the condition is satisfied and 0 otherwise. Signals with higher \emph{MuxBranch score} are more likely to influence conditional execution paths and multi-way branching behaviors.

\paragraph{\textbf{Hybrid Importance Score}}
All four feature metrics are normalized using min-max scaling to ensure comparability. The final hybrid importance score $Score(v)$ of a signal $v$ is computed as a weighted sum:

\begin{equation}
\begin{split}
Score(v) = & \; \alpha \cdot PR(v) + \beta \cdot Obs(v) \\
           & + \gamma \cdot OutBoost(v) + \delta \cdot MuxBranch(v)
\end{split}
\end{equation}
where the weighting coefficients are empirically set to $\alpha=0.45$, $\beta=0.25$, $\gamma=0.20$, and $\delta=0.10$ based on experimental evaluation, balancing structural centrality, backward influence, output relevance, and branching control to produce a comprehensive measure of signal importance for guiding subsequent RTL slicing and LLM-based assertion generation.

We evaluated several candidate features and weighting configurations, finding that the reported configuration achieved strong performance in identifying functionally important signals. Since no ground-truth labeling of signal importance is available, and commercial formal tools (e.g., JasperGold) do not expose internal relevance scores, we relied on manual inspection of generated assertions to assess configuration quality. Importantly, we intentionally limited parameter tuning to a single Block-level and a single CPU-level benchmark; once this configuration demonstrated stable and interpretable results, we froze the weights and applied them uniformly to all remaining designs without further adjustment.   This practice ensures that our reported accuracy reflects genuine generalization rather than overfitting to per-design idiosyncrasies.

\subsubsection{\textbf{Top-K Signal Selection}}

After scoring, we perform redundancy-aware Top-K selection to avoid choosing structurally equivalent signals, aiming to identify a compact yet expressive set of signals that maximizes code coverage while minimizing redundancy.

\paragraph{\textbf{Bidirectional Reachability-Based Similarity}}

For any signal node $u$, we define its forward and backward reachable sets as:
\begin{equation}
\begin{aligned}
\mathcal{R}^+(u) &= \text{reachable successors of } u, \\
\mathcal{R}^-(u) &= \text{reachable predecessors of } u
\end{aligned}
\end{equation}

Based on these definitions, we compute the Jaccard \cite{21} similarity over forward and backward reachable sets, respectively:
\begin{equation}
\begin{aligned}
J^+(u,v) &= \frac{|\mathcal{R}^+(u) \cap \mathcal{R}^+(v)|}{|\mathcal{R}^+(u) \cup \mathcal{R}^+(v)|}, \\
J^-(u,v) &= \frac{|\mathcal{R}^-(u) \cap \mathcal{R}^-(v)|}{|\mathcal{R}^-(u) \cup \mathcal{R}^-(v)|}
\end{aligned}
\end{equation}

The overall bidirectional Jaccard similarity between signal pairs is defined as a weighted combination of forward and backward similarities:
\begin{equation}
J(u,v) = \lambda \cdot J^+(u,v) + (1-\lambda) \cdot J^-(u,v)
\end{equation}
where $\lambda$ is an empirically assigned weighting coefficient that balances the contributions of forward and backward similarities. In our experiments, $\lambda$ is set to $0.5$.

\paragraph{\textbf{Greedy Selection with Priority}}

Signals are first sorted in descending order according to the hybrid score and then processed in a greedy manner. For each candidate signal, if its bidirectional Jaccard similarity with any previously selected signal exceeds the threshold $\theta = 0.4$, supported by the optimal partitioning bound for multi-set Jaccard similarity~\cite{28}, the signal with the lower score will be discarded; otherwise, the candidate is added to the selected set. This process iterates over the entire ranked signal list until a final filtered set of representative signals is obtained and output.

It is particularly noteworthy that purely combinational internal signals identified during RTL parsing are proactively excluded at this stage, as they generally do not contribute significantly to multi-cycle functional behaviors and may introduce noise into the signal ranking. We finally output the Top-K signal list, with the module each signal belongs to, as well as its corresponding category.

\subsection{Assertion Generation}

\subsubsection{\textbf{Structure-Aware RTL Slicing}}

To facilitate targeted assertion generation, we first perform structure-aware RTL slicing based on the extracted Top-K signals. Each selected signal serves as a slicing root, and the RTL semantic graph is traversed backwards to include all dependent nodes. Both temporal and control dependencies are considered, ensuring that the slice preserves relevant combinational and sequential logic paths. This approach significantly alleviates the limitations of prior methods, which feed the full RTL code directly into the model, often causing token overflow or making it difficult to capture critical behaviors when dealing with large-scale circuits due to excessively long contexts, thereby rendering them inapplicable to large-scale circuits.

\subsubsection{\textbf{Targeted Assertion Generation}}

\begin{figure}
\centering
\includegraphics[width=1\linewidth]{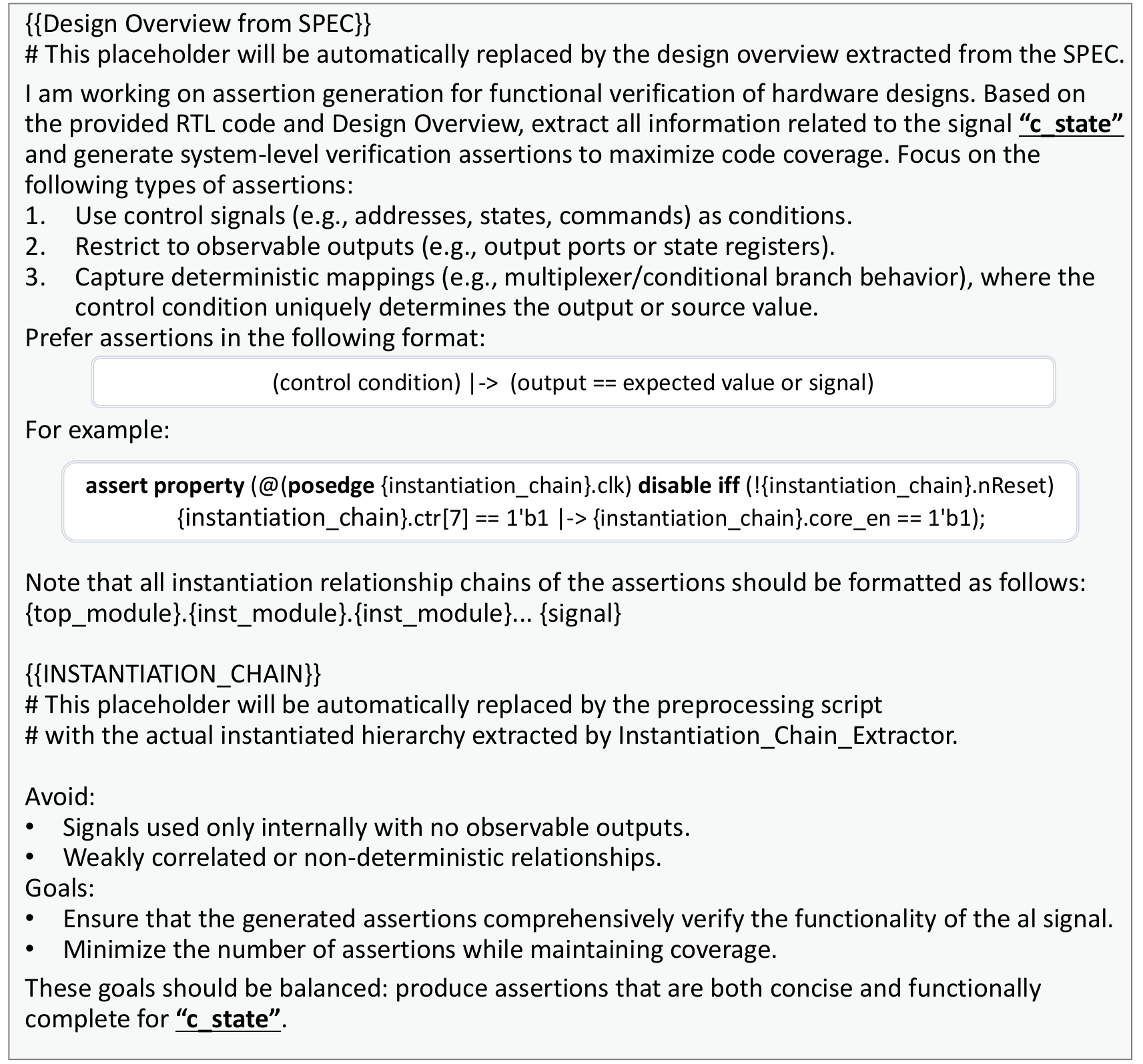}
\caption{Example prompt for targeted assertion generation based on RTL slices of the \texttt{c\_state} signal in the \texttt{i2c\_master\_byte\_ctrl} module.}
\label{fig:prompt}
\end{figure}

\begin{figure}[h]
\centering
\includegraphics[width=1\linewidth]{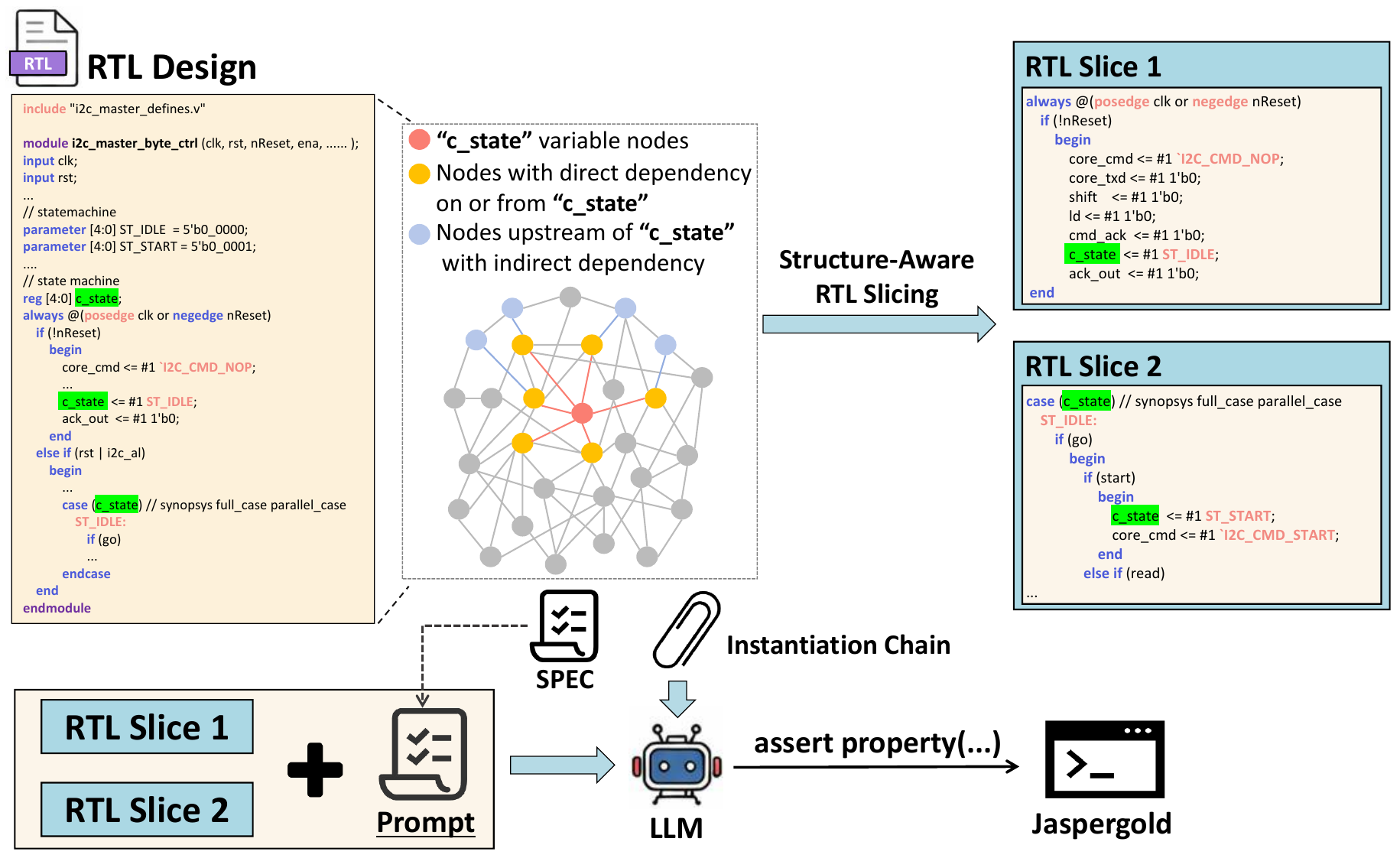}
\caption{Signal-centric RTL slicing and prompt-driven assertion generation: an example with the \texttt{c\_state} signal in the I2C design.}
\label{fig:generation}
\end{figure}

Once the RTL slices are extracted, assertions are generated for each signal by leveraging an LLM. As shown in Fig.~\ref{fig:prompt}, we design a prompt for this task and combine it with the overall design overview extracted from the specification file, the corresponding RTL slices of each signal, and the hierarchical module instantiation chain extracted by the parser (e.g., \texttt{\{top\_module\}.\{inst\_module\}.\{inst\_module\}...}) to guide the LLM in producing targeted assertions. For each generation round, only the assertions that are verified by Jaspergold \cite{26} as correct are retained for metric evaluation; if no correct assertion is obtained within three attempts, the signal is skipped. The overall workflow is illustrated in Fig.\ref{fig:generation}.

It is important to note that the correctness of the generated assertions is not the primary focus of this framework.  Our main goal is to identify critical signals for efficient verification of circuits.  Since some signal behaviors are inherently difficult for LLMs to interpret, assertion generation correctness can be low.  Therefore, we concentrate on evaluating the impact of correctly generated assertions for the selected signals on coverage metrics, which serves to validate our hypothesis. Even if the LLM repeatedly produces incorrect assertions, the framework remains meaningful, as engineers can still manually create targeted assertions for these critical signals in practical scenarios. This targeted approach ensures that the generated or manually crafted assertions are both non-redundant and coverage-efficient, focusing verification effort on signals that are most likely to impact design correctness. However, to ensure fair comparison with other similar methods and to explore how this framework addresses LLM context overflow in large-scale circuits, we still conduct LLM-based experiments to maintain a consistent experimental environment.

\section{Experiments}

\subsection{Benchmarks and Experimental Setup}

To intuitively assess AgileAssert, we select benchmark designs from both block- and CPU-level datasets, covering different abstraction levels and scales. A primary source is the dataset in \cite{23}, comprising 20 designs, each with a specification file and golden RTL code, mainly individual modules of varying scales. We also include three additional datasets \cite{24, 25, 29} of complete CPU designs, generally more complex, to evaluate our framework’s effectiveness and scalability on large-scale systems. Table~\ref{Summary of Benchmarks} details the designs, with LoC denoting post-synthesis lines of code.

\begin{table}[h]
\centering
\renewcommand{\arraystretch}{1.2}
\caption{Summary of benchmarks.}
\label{Summary of Benchmarks}
\setlength{\tabcolsep}{4pt}
\begin{tabular}{c|c|c|c}
\hline
\hline
\cellcolor{gray!20}\textbf{Design Level} & 
\cellcolor{gray!20}\textbf{Design Name} & 
\cellcolor{gray!20}\textbf{LoC} &
\cellcolor{gray!20}\textbf{Num. of Cells} \\ \hline

\multirow{3}{*}{Block-Level} 
& \texttt{I\textsuperscript{2}C} & 5369 & 756\\
& \texttt{SHA3} & 141185 & 22228\\  
& \texttt{Pairing} & 1561498 & 228287\\ 

\hline
\hline

\multirow{3}{*}{CPU-Level}  
& \texttt{picorv32} & 19651 & 9587\\  
& \texttt{tinyriscv} & 2435083 & 582488\\ 
& \texttt{e203} & 33644956 & 3323820\\

\hline
\hline
\end{tabular}
\end{table}

Correctness is analyzed using Cadence JasperGold (v21.12.002). All experiments are conducted on a server equipped with an Intel Xeon Gold 6148 CPU at 2.40GHz. To evaluate the effectiveness of AgileAssert, we perform comparative experiments against three state-of-the-art assertion generation methods under uniform conditions, with all methods using GPT-5.1 as the underlying model.

\begin{itemize}
    \item \textbf{AssertLLM \cite{6}:} Generates SVAs from specifications and waveform descriptions using multiple specialized LLMs, capturing bit-width and functional properties.
    \item \textbf{AssertGen \cite{11}:} Constructs cross-layer signal chains from high-level specifications to RTL, enabling RTL-stage assertions covering top- and module-level signals.
    \item \textbf{AssertMiner \cite{12}:} Uses LLMs to extract module-level specifications and fine-grained assertions from RTL, capturing micro-architectural behaviors.
\end{itemize}

We note that classical assertion mining techniques (e.g., GoldMine) have been systematically evaluated and shown to produce assertions with significantly weaker verification effectiveness than LLM-based approaches in prior work \cite{6,8}. Their reliance on simulation traces or fixed templates prevents them from capturing complex design intent or deep hierarchical dependencies. Consequently, we focus our comparison only on LLM-based methods, which represent the current state of the art and the most relevant baselines for our work.

\subsection{Evaluation Metrics}

We adopt a set of evaluation metrics provided by Cadence JasperGold (v21.12.002) \cite{26}, as summarized in Table~\ref{Summary of Evaluation Metrics}. These metrics assess the effectiveness of the generated assertions from multiple perspectives, including coverage, proof quality, and fault detection capability.

\begin{table}[h]
\centering
\renewcommand{\arraystretch}{1.2}
\caption{Summary of evaluation metrics.}
\label{Summary of Evaluation Metrics}
\begin{tabular}{c|>{\raggedright\arraybackslash}p{7cm}} 
\hline
\hline
\cellcolor{gray!20}\textbf{Metrics} & \multicolumn{1}{c}{\cellcolor{gray!20}\textbf{Description}} \\ 
\hline
$N$ & Number of generated correct SVAs \\ 
\textbf{$BFC$} & Branch coverage within the COI of generated assertions \\ 
\textbf{$SFC$} & Statement coverage within the COI of generated assertions \\ 
\textbf{$TFC$} & Toggle coverage within the COI of generated assertions \\
\textbf{$COI$} & Cone of influence covered by assertions \\
\textbf{$ER$} & Error detection rate in mutation testing \\
\hline
\hline
\end{tabular}
\end{table}

\definecolor{deepgreen}{RGB}{110, 251, 152}
\definecolor{lightgreen}{RGB}{188, 251, 152}

\begin{table*}
\begin{adjustwidth}{-0.35cm}{0cm}  
\centering
\setlength{\tabcolsep}{4.5pt} 
\renewcommand{\arraystretch}{1} 
\caption{Coverage metrics for different designs with varying number of selected signals}
\label{tab:experimental_1}
\begin{tabular}{c ccc ccc ccc ccc ccc ccc}
\toprule
 & \multicolumn{3}{c}{\textbf{I2C}} & \multicolumn{3}{c}{\textbf{SHA3}} & \multicolumn{3}{c}{\textbf{Pairing}} & \multicolumn{3}{c}{\textbf{picorv32}} & \multicolumn{3}{c}{\textbf{tinyriscv}} & \multicolumn{3}{c}{\textbf{e203}} \\
\cmidrule(lr){2-4} \cmidrule(lr){5-7} \cmidrule(lr){8-10} \cmidrule(lr){11-13} \cmidrule(lr){14-16} \cmidrule(lr){17-19}
\textbf{\#signal} & 1 & 3 & 5 & 1 & 3 & 5 & 1 & 3 & 5 & 1 & 3 & 5 & 1 & 3 & 5 & 1 & 3 & 5 \\
\midrule
\textbf{N} & 5 & 15 & 28 & 4 & 16 & 36 & 1 & 9 & 15 & 1 & 4 & 7 & 4 & 15 & 21 & 2 & 4 & 8 \\
\textbf{BFC* (\%)} & 86.07 & \cellcolor{lightgreen}87.71 & 87.71 & 64 & 72 & \cellcolor{lightgreen}88 & \cellcolor{lightgreen}93.85 & 93.85 & 93.85 & 34.08 & \cellcolor{lightgreen}34.71 & 34.71 & 85.37 & \cellcolor{lightgreen}85.85 & 85.85 & 48.16 & 48.16 & \cellcolor{lightgreen}48.51 \\
\textbf{SFC* (\%)} & 84.27 & \cellcolor{lightgreen}89.52 & 89.52 & 43.90 & 48.78 & \cellcolor{lightgreen}87.81 & \cellcolor{lightgreen}94.67 & 94.67 & 94.67 & 30.97 & \cellcolor{lightgreen}32.26 & 32.26 & 83.77 & \cellcolor{lightgreen}84.15 & 84.15 & 42.62 & 42.62 & \cellcolor{lightgreen}48.33 \\
\textbf{TFC* (\%)} & 80.14 & \cellcolor{lightgreen}94.77 & 94.77 & 2.97 & 2.98 & \cellcolor{lightgreen}78.18 & \cellcolor{lightgreen}81.19 & 81.19 & 81.19 & \cellcolor{lightgreen}23.80 & 23.80 & 23.80 & 69.87 & 70.27 & \cellcolor{lightgreen}70.83 & 16.32 & 16.32 & \cellcolor{lightgreen}18.66 \\
\textbf{COI* (\%)} & 87.59 & 97.31 & \cellcolor{lightgreen}97.46 & 3.4 & 3.45 & \cellcolor{lightgreen}78.34 & 99.51 & 99.51 & \cellcolor{lightgreen}99.52 & 50.43 & \cellcolor{lightgreen}51.92 & 51.92 & 86.55 & \cellcolor{lightgreen}86.61 & 86.61 & 69.28 & 69.28 & \cellcolor{lightgreen}73.19 \\
\textbf{ER (\%)} & 1.59 & 9.26 & \cellcolor{lightgreen}13.49 & 10.23 & 19.32 & \cellcolor{lightgreen}23.86 & 0.47 & 1.40 & \cellcolor{lightgreen}2.48 & 0 & 6.10 & \cellcolor{lightgreen}7.02 & 0.95 & 2.15 & \cellcolor{lightgreen}2.65 & 0.31 & 0.77 & \cellcolor{lightgreen}0.93 \\
\bottomrule
\end{tabular}
\end{adjustwidth}

\vspace{0.1cm}
\scriptsize
\raggedright
\hspace{3em}*Note:\hspace{0.5em}\textbf{\#signal} indicates the number of signals selected from the top-k list for generating assertions. Cells highlighted in light green mark the points at which each metric \\
\hspace{4em}reaches its maximum for a given design.

\end{table*}

\begin{table*}[ht]
\begin{adjustwidth}{-0.6cm}{0cm}  
\setlength{\tabcolsep}{1pt} 
\renewcommand{\arraystretch}{1} 
\caption{Beating baselines with minimal assertions for superior coverage metrics via stepwise signal addition}
\label{tab:baseline_comparison}
\footnotesize
\begin{tabular}{c cccc cccc cccc c}
\toprule
\multirow{2}{*}{\textbf{Block-Level}} 
& \multicolumn{4}{c}{\textbf{I2C}} 
& \multicolumn{4}{c}{\textbf{SHA3}} 
& \multicolumn{4}{c}{\textbf{Pairing}} & \multirow{2}{*}{\textbf{Avg.}} \\
\cmidrule(lr){2-5} \cmidrule(lr){6-9} \cmidrule(lr){10-13}
 & AssertLLM & AssertGen & AssertMiner & \textbf{Ours (\#3)} 
 & AssertLLM & AssertGen & AssertMiner & \textbf{Ours (\#6)} 
 & AssertLLM & AssertGen & AssertMiner & \textbf{Ours (\#1)} &  \\
\midrule
\textbf{N} & 64 & 86 & 37 & 15 & 44 & 42 & 40 & 39 & 32 & 67 & 27 & 1 & \textbf{\hspace{1em}$\downarrow$ 59.03\%} \\
\textbf{BFC* (\%)} & 81.15 & 82.61 & 82.79 & \cellcolor{lightgreen}87.71 & 92 & 92 & 88 & \cellcolor{lightgreen}100 & 76.12 & 79.61 & 89.82 & \cellcolor{lightgreen}93.85 & \textbf{\hspace{-0.5em}$\uparrow$ 8.99} \\
\textbf{SFC* (\%)} & 82.26 & 82.26 & 83.09 & \cellcolor{lightgreen}89.52 & 90.24 & 90.24 & 90.84 & \cellcolor{lightgreen}95.12 & 83.64 & 72.27 & 88.66 & \cellcolor{lightgreen}94.67 & \textbf{\hspace{-0.5em}$\uparrow$ 8.27} \\
\textbf{TFC* (\%)} & 83.08 & 79.01 & 79.62 & \cellcolor{lightgreen}94.77 & 63.91 & 76.15 & 66.04 & \cellcolor{lightgreen}78.26 & 70.93 & 72.77 & 57.64 & \cellcolor{lightgreen}81.19 & \textbf{$\uparrow$ 12.61} \\
\textbf{COI* (\%)} & 81.65 & 90.8 & 91.94 & \cellcolor{lightgreen}97.31 & 82.91 & \cellcolor{lightgreen}83.01 & 80.98 & 78.52 & 30.04 & 48.64 & 67.02 & \cellcolor{lightgreen}99.51 & \textbf{$\uparrow$ 18.78} \\
\midrule

\multicolumn{14}{c}{%
\makebox[\textwidth][c]{%
\begin{tabular}{c cccc cccc cccc c}
\multirow{2}{*}{\textbf{CPU-Level}} 
 & \multicolumn{4}{c}{\textbf{picorv32}} 
 & \multicolumn{4}{c}{\textbf{tinyriscv}} 
 & \multicolumn{4}{c}{\textbf{e203}} & \multirow{2}{*}{\textbf{Avg.}} \\
\cmidrule(lr){2-5} \cmidrule(lr){6-9} \cmidrule(lr){10-13}
 & AssertLLM & AssertGen & AssertMiner & \textbf{Ours (\#3)} 
 & AssertLLM & AssertGen & AssertMiner & \textbf{Ours (\#1)} 
 & AssertLLM & AssertGen & AssertMiner & \textbf{Ours (\#1)} &  \\
 \midrule
\textbf{N}   & 37 & 43 & 6 & 4 & 25 & 37 & 7 & 4 & 9 & 16 & \textbackslash & 2 & \textbf{\hspace{1.3em}$\downarrow$ 74.32\%} \\
\textbf{BFC* (\%)} & 18.47 & 19.75 & \cellcolor{lightgreen}34.71 & \cellcolor{lightgreen}34.71 & 61.18 & 66.64 & 81.26 & \cellcolor{lightgreen}85.37 & 0 & 21.09 & \textbackslash & \cellcolor{lightgreen}48.51 & \textbf{\hspace{0.3em}$\uparrow$ 19.27} \\
\textbf{SFC* (\%)} & 23.06 & 24.84 & \cellcolor{lightgreen}32.26 & \cellcolor{lightgreen}32.26 & 60.89 & 60.94 & 80.62 & \cellcolor{lightgreen}83.77 & 0 & 20.91 & \textbackslash & \cellcolor{lightgreen}48.33 & \textbf{\hspace{0.3em}$\uparrow$ 17.65} \\
\textbf{TFC* (\%)} & 14.21 & 16.42 & 23.56 & \cellcolor{lightgreen}23.80 & 45.35 & 52.95 & 64.34 & \cellcolor{lightgreen}69.87 & 0 & 10.54 & \textbackslash & \cellcolor{lightgreen}18.66 & \textbf{\hspace{0.3em}$\uparrow$ 11.37} \\
\textbf{COI* (\%)} & 30.89 & 44.57 & \cellcolor{lightgreen}51.92 & \cellcolor{lightgreen}51.92 & 55.13 & 68.89 & 86.43 & \cellcolor{lightgreen}86.55 & 0 & 33.85 & \textbackslash & \cellcolor{lightgreen}73.19 & \textbf{\hspace{0.3em}$\uparrow$ 23.76} \\
\end{tabular}%
}} \\
\bottomrule

\end{tabular}
\end{adjustwidth}
\vspace{0.1cm}
\scriptsize
\raggedright
*Note:\hspace{0.5em}\textbf{\#n} indicates the number of signals selected from the top-k list for generating assertions. The cells highlighted in light green indicate the cells where each metric reaches its maximum value under the specific design. \textbf{``\textbackslash''} indicates a generation failure caused by the input token length exceeding the model's maximum token limit.

\end{table*}

However, we notice that these metrics are computed based on the Cone of Influence (COI \cite{26}) covered by the input assertions. As different assertion sets may involve different signals (i.e., different denominators), direct comparison across experiments will be unfair.

To enable fair comparison among different assertion sets within the same experimental batch, we propose a set of normalized metrics, denoted as \textbf{$BFC^*$}, \textbf{$SFC^*$}, \textbf{$TFC^*$} and \textbf{$COI^*$}. These metrics are computed by using the largest COI observed in the same batch as the denominator. Formally, let $M \in \{BFC, SFC, TFC, COI\}$ be a coverage metric and let $\text{Cov}(M)$ denote the number of items actually covered by the assertions in the current experiment. Let $\text{Num}_{\max}^M$ denote the largest number of items relevant to metric $M$ among all assertion sets in the same batch. Then the normalized metric $M^*$ is defined as:
\[
M^* = \frac{\text{Cov}(M)}{\text{Num}_{\max}^M}
\]

It should be noted that these normalized metrics still do not reflect the total verification coverage of the entire design, because JasperGold can not obtain the complete \textbf{$COI$} of the full design. Nevertheless, they enable fairer comparison across different assertion sets within the same batch.

\subsection{Experimental Results}

\subsubsection{\textbf{Incremental Selection of Top-K Signals}}

We evaluate the coverage achieved by incrementally selecting signals from the top-K list for each design. For each circuit, we select the top 5 signals according to their ranking. Assertions are generated incrementally: each new signal adds assertions that are combined with the previously generated ones, and the resulting cumulative coverage is measured. 

Table~\ref{tab:experimental_1} reports the coverage obtained after selecting 1, 3, and 5 signals, which is particularly encouraging. As Table~\ref{tab:experimental_1} shows, for most block-level designs, generating assertions for only the top-ranked signal achieves remarkably high coverage. Even for the large-scale Pairing design, a single top-ranked assertion achieves a \textbf{$COI^*$} of 99.51, highlighting this signal's hub role. Each additional assertion further improves the error detection rate (\textbf{$ER$}), a metric demanding strong behavioral constraints, demonstrating the high quality and effectiveness of the generated assertions with targeted signals.

For SHA3, top-ranked signals initially contribute little to coverage; only when the fourth signal is included does coverage rise significantly. Closer examination shows that SHA3 exhibits a highly irregular structure: the code contains numerous generated constructs that replicate logic multiple times, causing a signal present only once in the source code to appear hundreds of times after expansion. Our ranking model does not account for such expansions, indicating potential room for more optimal signal ranking.

Furthermore, although the generation accuracy of assertions is not the primary focus of this work, it is worth noting that almost all the aforementioned assertions were captured within a single generation round. For Blcok-level designs, all assertions generated from the top five signals achieved an accuracy above \textbf{80\%}, while for CPU-level designs, the accuracy exceeded \textbf{60\%}, which is also far more than the other three comparison methods. We give the specific generated data for each circuit in Table~\ref{tab:agileassert_accuracy}.

\begin{table}[htbp]
\centering
\caption{Assertion generation accuracy of AgileAssert}
\label{tab:agileassert_accuracy}
\begin{tabular}{c|ccc|ccc}
\hline
\multirow{2}{*}{\textbf{AgileAssert}} & \multicolumn{3}{c|}{\textbf{Block-Level}} & \multicolumn{3}{c}{\textbf{CPU-Level}} \\
\cline{2-7}
 & \textbf{I2C} & \textbf{SHA3} & \textbf{Pairing} & \textbf{picorv32} & \textbf{tinyriscv} & \textbf{e203} \\
\hline
$\mathbf{N/N_g}$ & 28/33 & 36/44 & 15/15 & 7/14 & 21/23 & 8/13 \\
\hline
\textbf{Avg.} & \multicolumn{3}{c|}{88.89\%} & \multicolumn{3}{c}{67.61\%} \\
\hline
\end{tabular}
\\[6pt]
\scriptsize
\raggedright
*Note:\hspace{0.2em} $N_g$ represents the total number of generated assertions, while $N$ represents the number of assertions that have been verified to be correct through formal validation. \textit{Avg.} represents the average generation accuracy rate of each type of circuit.
\end{table}

The above results were all generated by the large language model in one round. In our subsequent attempts, we found that these results were sufficiently stable and even achieved better outcomes than those in the manuscript. This high accuracy can be attributed to our ability to provide precise assertion targets and targeted contextual code snippets for generation.

\subsubsection{\textbf{Stepwise Signal Addition to Boost Coverage Metrics Beyond Baseline}}

To evaluate the proposed signal ranking strategy, we conduct a stepwise signal addition experiment. Starting from the top-ranked signal, signals are incrementally added from the ranked list to generate assertions until either all coverage metrics surpass those of the baseline methods, or the assertion count is about to exceed that of other methods. This setup evaluates both the effectiveness of the selected signals and how efficiently high coverage can be achieved with minimal assertions. The results are summarized in Table~\ref{tab:baseline_comparison}. For most circuits, assertions generated from only the top three signals already outperform all baselines across all metrics while requiring significantly fewer assertions. Even for the large-scale Pairing and e203 designs, our method surpasses all baselines on every metric using only one and two assertions, respectively. For the most complex and largest-scale e203 design, other methods either fail due to token overflow or generate assertions with very low accuracy; in particular, AssertLLM produces correct assertions only for bit-width checking, which do not contribute to coverage. For SHA3, due to the structural peculiarities mentioned above, the performance is slightly worse than other methods only on the \textbf{$COI*$} metric when using a small number of assertions.

\begin{figure}[h]
\centering
\includegraphics[width=1\linewidth]{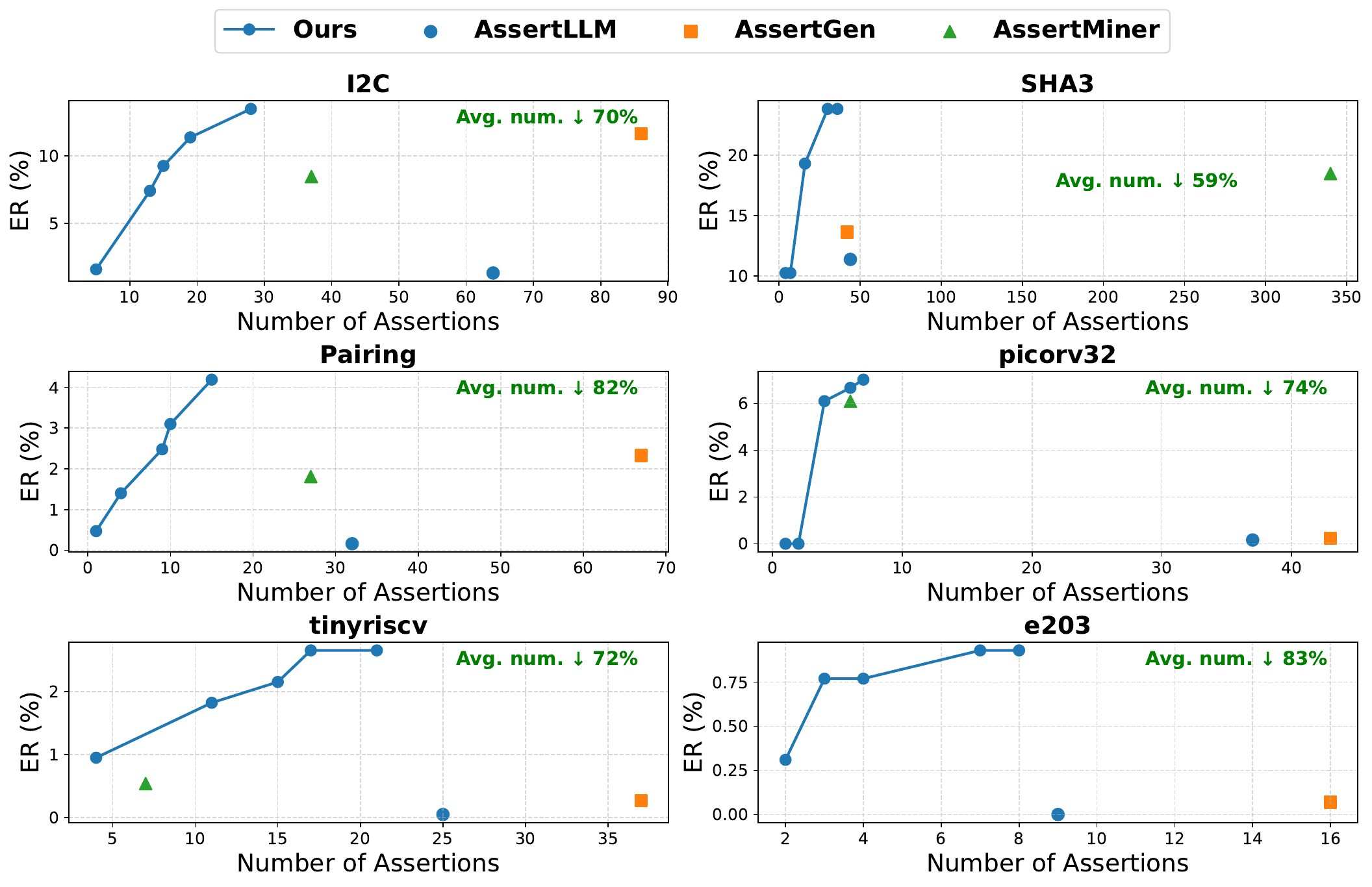}
\caption{Progressive error detection rate improvement via stepwise signal addition compared to baseline methods.}
\label{fig:mutation}
\end{figure}

Fig.~\ref{fig:mutation} illustrates mutation testing results. Baselines are shown as fixed points corresponding to their final error detection rates (\textbf{$ER$}), while our method is represented as a progressive curve, eventually surpassing all baselines with an average \textbf{72.74\%} reduction in the number of assertions when compared to other methods.

Analysis shows that critical signals identified by our approach are often internal to lower-level modules. Most previous methods rarely generate assertions for such signals, since specifications provide little or no information about them. While AssertMiner can produce assertions involving these signals, its guidance does not have a focus. In contrast, our approach explicitly centers on critical signals, allowing us to provide signal-focused, precise guidance to the LLM. By supplying RTL slices and context specific to each signal, we ensure that the generated assertions directly monitor the behavior of high-impact signals. This targeted, signal-driven strategy results in assertions that are both highly effective and strongly constraining, outperforming prior methods in verification quality.

These results provide strong evidence supporting the validity of our approach: \textbf{identifying and focusing on the most critical signals in a circuit is the key to effective verification}. By targeting high-utility signals, our method not only achieves high coverage with a minimal number of assertions but also demonstrates that a more selective, signal-driven strategy can outperform approaches that indiscriminately generate a large number of assertions.

\subsubsection{\textbf{Efficiency Analysis: Token Cost and Verification Coverage}}

\begin{figure}[h]
\centering
\includegraphics[width=0.9\linewidth]{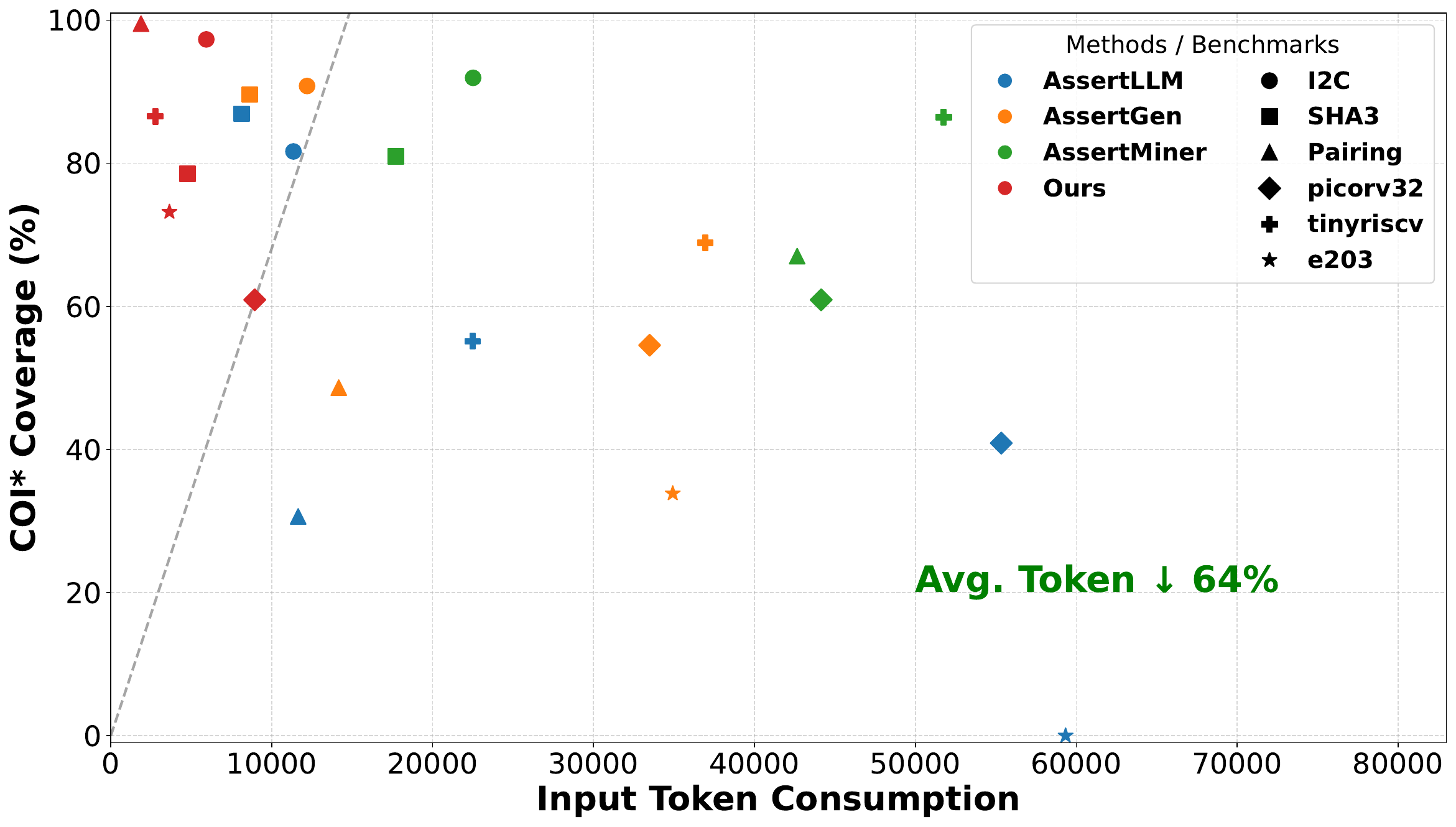}
\caption{COI coverage vs. input token consumption across benchmarks.}
\label{fig:token}
\end{figure}

To evaluate the cost-effectiveness of assertion generation, we compare input token consumption against the achieved verification coverage. Among all metrics, COI most directly reflects the verification scope and is therefore adopted as the primary metric for efficiency comparison. The corresponding results are presented in Fig.~\ref{fig:token}.

The figure shows that our method achieves higher COI coverage with an average \textbf{64\%} reduction in token consumption compared to other approaches, demonstrating its efficiency. This indicates that targeting critical signals enables more cost-effective verification without sacrificing coverage.

\subsubsection{\textbf{A Case Study: Why AgileAssert Substantially Outperforms Baselines}}

\begin{figure}[h]
\centering
\includegraphics[width=1\linewidth]{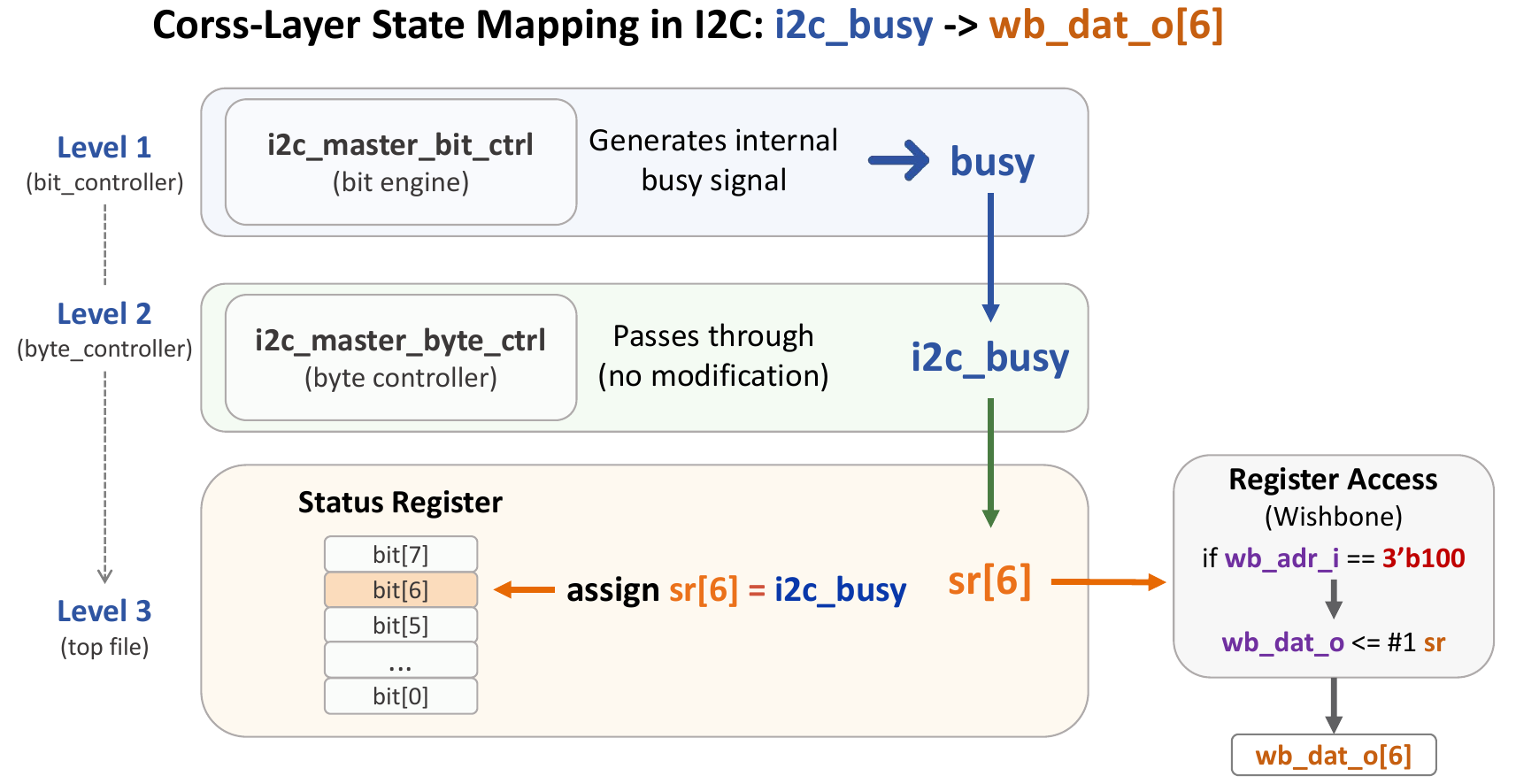}
\caption{A three-level hierarchical dependency captured by AgileAssert in the I2C benchmark.}
\label{fig:case_study}
\end{figure}

To concretely illustrate the advantage of our signal-driven approach, we examine a representative assertion generated for the I2C benchmark:

\begin{lstlisting}[language=Verilog]
assert property (@(posedge wb_clk_i) disable iff (arst_i ^ ARST_LVL)
    (wb_adr_i == 3'b100) |-> ##1 wb_dat_o[6] == i2c_busy);
\end{lstlisting}

This assertion achieves high coverage, because it precisely captures a cross-layer state-mapping invariant: when address 3'b100 selects the status register, bit 6 of the Wishbone data output must reflect the internal \texttt{i2c\_busy} signal one cycle later — a three-level hierarchical dependency that is virtually impossible to trace correctly by other methods without crtical signal analysis, as shown in Fig.~\ref{fig:case_study}.

\textbf{Why traditional methods fail.} Classical assertion mining techniques, such as trace-based or template-driven approaches (e.g., GoldMine), rely on simulation waveforms or pre-defined templates. Without structural signal analysis, they cannot trace deep hierarchical dependencies across module boundaries, nor can they infer the semantic relationship between a register address and an internal status flag. Consequently, they either miss such invariants entirely or produce superficial assertions with weak verification value, as systematically analyzed in prior work \cite{6,8}.

\textbf{Why other LLM-based methods fail.} Existing LLM-based approaches typically prompt the model with raw RTL or specifications without identifying functionally critical signals first.   This leads to two fundamental problems.   First, the LLM may target arbitrary, non-essential signals with low verification value, or fail to discover deep, critical signals buried in internal modules without structural guidance.   Second, even when the LLM happens to select a critical signal, the absence of precise slicing floods it with excessive or irrelevant context, causing it to generate superficial assertions—e.g.  , checking signal stability under generic conditions or correlating signals without functional semantics—rather than precise temporal invariants.

\textbf{Why AgileAssert succeeds.} Our framework first identifies \texttt{i2c\_busy} as a structurally central and verification-relevant signal via hybrid ranking, then slices the RTL to expose exactly the relevant context: the register decoding logic, the status register definition, and the internal busy flag generation. This targeted guidance enables the LLM to infer the correct temporal relationship ($\mid\rightarrow\#\#1$) and data-path invariant (\texttt{wb\_dat\_o[6]} == \texttt{i2c\_busy}) with explicit verification intent. Without explicitly targeting the key signal and feeding precisely sliced fragments, capturing such a deep, cross-module dependency is virtually impossible. This case exemplifies why our signal-driven paradigm achieves substantially higher coverage and formal verifiability than both traditional tools and unconstrained LLM-based baselines.

\subsubsection{\textbf{Ablation Study}}

\begin{figure}[h]
\centering
\includegraphics[width=0.8\linewidth]{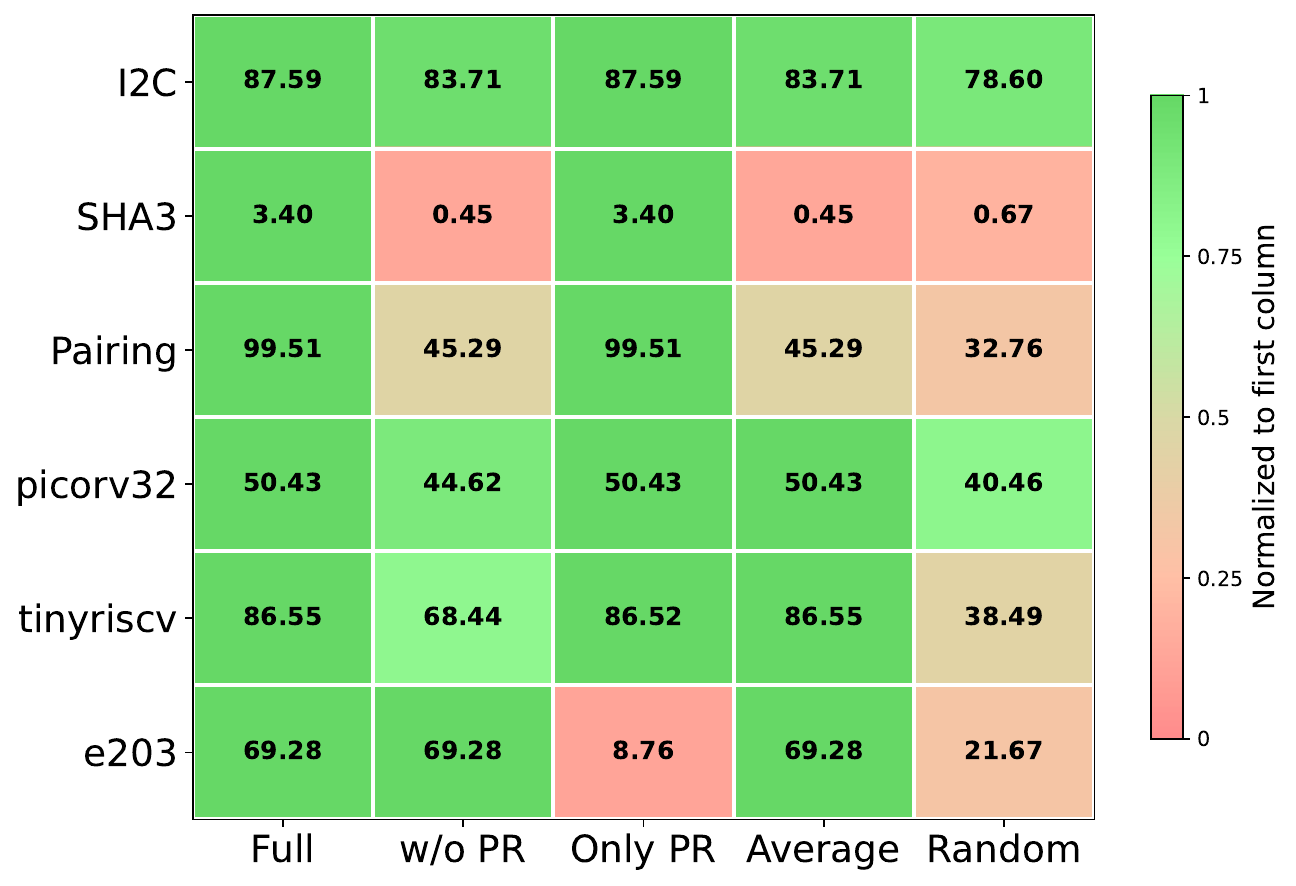}
\caption{Ablation study on \textit{COI*} coverage using different signal ranking configurations.}
\label{fig:ablation}
\end{figure}

To validate the proposed signal ranking strategy, we conduct ablation experiments comparing five configurations: complete hybrid scoring (\textit{Full}), without PageRank (\textit{w/o PR}), PageRank only (\textit{Only PR}), equal weighting of four features (\textit{Average}), and random signal ranking (\textit{Random}). We evaluate only the top-ranked signal, as it efficiently validates the ranking strategy—correctly prioritizing the most important signal shows model effectiveness. Among the configurations, PageRank dominates the top-ranked signals, while other feature scores have smaller coefficients and removing any one of them does not noticeably affect the top-ranked signals. Bidirectional Jaccard, important for signal complementarity, is not included as it does not affect the top-ranked signal. If the top-ranked signal has been evaluated in previous experiments, we directly reuse its previously generated assertion results to mitigate LLM randomness and ensure consistency. Fig.~\ref{fig:ablation} shows the \textbf{$COI^*$} coverage for each configuration across six benchmark designs.

Fig.~\ref{fig:ablation} shows that the \textit{Full} configuration consistently achieves the highest \textbf{$COI^*$} coverage. The \textit{Only PR}, \textit{w/o PR} and \textit{Average} configurations perform close to \textit{Full} in most designs, but exhibit noticeable drops in a few cases, demonstrating the stability of our framework's performance. In contrast, \textit{Random} shows significant decreases across most designs, highlighting the importance of the proposed hybrid ranking strategy. 

Notably, we observed that \textit{Only PR} performs comparably to the Full configuration on most designs, with a pronounced drop only on e203. This aligns with the nature of PageRank as a topology-centric metric: in small or structurally simple designs, structurally central signals tend to coincide with functionally critical and observable ones, making structural ranking alone sufficient. However, e203 is a significantly more complex CPU-level design with richer signal interactions and deeper dependencies. Here, relying solely on structural centrality risks over-prioritizing graph-central yet less informative signals—such as repetitive pipeline control or bus arbitration signals—thereby degrading assertion quality. This drop is not incidental but reflects an inherent limitation of pure structural metrics: they cannot distinguish between "structurally central" and "verification-relevant" signals. This further underscores the necessity of our hybrid ranking strategy, where observability and output relevance features effectively compensate for this blind spot, ensuring robust performance across diverse design complexities.

\subsubsection{\textbf{Practical Utility Despite the Absence of Golden RTL}}

Similar to other RTL-based approaches, assertion generation frameworks that operate directly on the design source code are inherently limited by the quality of the input RTL: if the design contains latent errors, assertions generated from its behavior may fail to detect these errors or, in the worst case, inadvertently encode the buggy functionality. This is a shared challenge of all techniques that operate without a formal golden model.

Nevertheless, as discussed earlier, a significant practical value of AgileAssert lies in its ability to directly pinpoint the signals most critical to functional verification, thereby assisting verification engineers even when assertions are written manually. By highlighting structurally central and verification-relevant signals, AgileAssert substantially reduces the manual effort required to identify assertion targets, accelerating the overall verification workflow.

What further distinguishes our framework is the nature of its assertion targets. By focusing on structurally stable architectural signals—such as module interfaces, protocol controls, and state-machine transitions—our method produces assertions that guard the design skeleton.   These interface-level constraints must satisfy protocols regardless of internal implementation errors, enabling the generated assertions to be reused across RTL iterations with minimal modification. This aligns with common industrial practices, where interface assertions are preserved and refined throughout the design cycle.

Thus, although AgileAssert cannot guarantee detection of arbitrary internal bugs when operating without a golden RTL, it offers a practical solution for generating—or assisting engineers in manually crafting—robust, reusable architectural assertions that remain valid across design iterations.

\section{Conclusion}

We introduce AgileAssert, a signal-centric LLM framework for efficient RTL verification. By identifying and prioritizing the most functionally critical signals, it generates a small set of highly targeted assertions that achieve superior coverage. Leveraging structure-aware RTL slicing and hybrid signal ranking, AgileAssert effectively addresses scalability challenges that have long plagued large-scale designs. Experiments on multiple open-source designs, ranging from block-level modules to million-gate CPUs, demonstrate significant improvements across diverse coverage metrics including branch, statement, toggle, and COI coverage, as well as error detection rates. These results validate that targeted, signal-driven verification is key to achieving efficient and scalable RTL validation.

\newpage
\bibliographystyle{plain} 
\bibliography{reference}

\end{document}